\documentclass{emulateapj}

\newcommand{\chandra}{{\it Chandra}}
\newcommand{\NH}{\mbox{$N_{\rm H}$}}        

\begin{document}

\title{The X-ray Halo of GX5-1} 
\author{Randall K. Smith}
\affil{Code 662, NASA/Goddard Space Flight Center, Greenbelt, MD 20771\\ 
Department of Physics and Astronomy, The Johns Hopkins University,
Baltimore, MD 21218, USA}
\email{rsmith@milkyway.gsfc.nasa.gov}

\author{T. M. Dame}
\affil{Harvard-Smithsonian Center for Astrophysics, 60 Garden St.,
Cambridge, MA 02138, USA}

\author{Elisa Costantini}
\affil{SRON National Institute for Space Research, Sorbonnelaan, 2,
  3584CA, Utrecht, The Netherlands \\
Astronomical Institute, Utrecht University, P.O. Box 80000, 3508TA
Utrecht, The Netherlands}

\author{Peter Predehl}
\affil{Max-Planck-Institut f\"ur extraterrestrische Physik,
  Giessenbachstr. 1, D-85748 Garching bei M\"unchen, Germany}

\begin{abstract}
Using \chandra\ observations we have measured the energy-resolved
dust-scattered X-ray halo around the low-mass X-ray binary GX5-1,
which shows signs of both singly- and multiply-scattered X-rays.  We
compared the observed X-ray halo at various energies to predictions
from a range of dust models.  These fits used both
smoothly-distributed dust as well as dust in clumped clouds, with CO
and 21~cm observations helping to determine the position of the clouds
along the line of sight.  We found that the BARE-GR-B model of
\citet{ZDA04} generally led to the best results, although inadequacies
in both the overall model and the data limit our conclusions.  We did
find that the composite dust models of \citet{ZDA04}, especially the
``no carbon'' models, gave uniformly poor results. Although models
using cloud positions and densities derived naively from CO and 21 cm
data gave generally poor results, plausible adjustments to the
distance of the largest cloud and the mass of a cloud in the expanding
3 kpc Arm lead to significantly improved fits.  We suggest that
combining X-ray halo, CO, and 21~cm observations will be a fruitful
method to improve our understanding of both the gas and dust phases of
the interstellar medium.
\end{abstract}

\keywords{dust --- scattering --- X-rays: binaries ---  X-rays: ISM }

\section{Introduction\label{sec:intro}}

Although interstellar (IS) dust is a significant component of dense
molecular clouds, measuring the properties of these grains is
difficult.  Except in the outskirts of the clouds, the inherently
large extinction prevents optical and UV measurements except in
unusual circumstances.  As a result, information about the size
distribution and composition of grains in dense clouds has by
necessity been extrapolated from observations of less-dense clouds or
modeled based on IR measurements.  We show that X-ray scattering
observations, combined with CO spectral line measurements, can put
useful limits on the allowed dust models.

Our X-ray source is the low-mass X-ray binary (LMXB) GX5-1 at ($l,b$) =
(5.077,-1.019).  GX5-1 is a Z source \citep[see][and references
  therein]{vdK95}, and is the second brightest persistent Galactic
X-ray source after the Crab Nebula.  The secondary has been imaged in
the IR \citep{Jonker00} but the IR spectra measured by \citet{Bandy03}
could not be typed.  \citet{Christian97} estimated the distance to be
9 kpc, although they warned that this is probably an upper limit to
the distance and assigned an error of 2.7 kpc to the value.  Using
X-ray spectra taken by the {\sl Einstein}\ satellite they also
measured the column density to be \NH $= 2.54 \pm 0.19 \times
10^{22}$\,cm$^{-2}$, although this result depends on the assumed
spectral model.  Recently, \citet{Ueda05} used a \chandra\ HETG
observation to measure the depth of absorption edges of a number of IS
metals along the line of sight to GX5-1.  Based on these measurements,
they determined the total \NH $=
2.8^{+3.3}_{-1.8}\times10^{22}$\,cm$^{-2}$; this result has larger
error bars than the \citet{Christian97} value, primarily because it
does not assume a spectral model {\it a priori}.

The X-ray halo around GX5-1 has been observed by many X-ray
satellites, albeit with lower angular or energy resolution than is
possible with the \chandra\ ACIS.  \citet{Predehl92} used a lunar
occultation of GX5-1 to measure the scattered halo directly, finding a
total halo intensity (in the ROSAT band) 28\% of the source intensity.
GX5-1 was also included in the \citet{PS95} survey of X-ray halos
detected with ROSAT, where they measured a halo intensity 25.4\% of
the source intensity.  Recently, \citet{Xiang05} analyzed 17 X-ray
sources, including GX5-1, using a new technique based on the
zero-order image of \chandra\ HETG data.  They found a lower total halo
intensity (9.9\%), as expected since the \chandra\ bandpass extends to
higher energies where halo intensities are smaller.  They also
measured a total column density \NH $= 2.0\times10^{22}$\,cm$^{-2}$\
and fit their data to determine the relative dust positions along the
line of sight, finding that 90\% of the dust was within 10\% of the
source.

\section{Theoretical Framework}

\citet{Overbeck65} first described astrophysical X-ray halos, which
result from small-angle scattering of X-rays inside solid matter.  The
technique has long been used in laboratory studies of virii or proteins
in solution, as careful measurements of the radial profile can show
the density and shape of the particles \citep{Guinier55}.  The theory
of X-ray halos in an astrophysical setting has been described in
detail by a number of authors \citep{MG86,ML91,SD98,Draine03}.  We briefly
review the theory here.

The fundamental quantity is the differential scattering cross section
$d\sigma/d\Omega$, which can be calculated using either the exact Mie
solution or the Rayleigh-Gans (RG) approximation; see \citet{SD98} for
a discussion.  As the RG approximation is analytic it demonstrates how
the various parameters scale:
\begin{eqnarray}
{{d\sigma(\theta_{\rm sca})}\over{d\Omega}} & = &
1.1\,\hbox{cm}^2\hbox{sr}^{-1} \Big({{2Z}\over{M}}\Big)^2
\Big({{\rho}\over{3 \hbox{g\,cm}^{-3}}} \Big)^2 a_{\mu\rm m}^6 \times
\\ \nonumber
 & & \Big({{F(E)}\over{Z}}\Big)^2 \Phi^2(\theta_{\rm sca})
\label{eq:RG}
\end{eqnarray}
where $a$\ is the grain radius, $Z$\ the mean atomic charge, $M$\ the
mean atomic weight (in amu), $\rho$\ the mass density, $E$\ the X-ray
energy in keV, $F(E)$ the atomic scattering factor \citep{Henke81},
$\theta_{\rm sca}$\ the scattering angle, and $\Phi^2(\theta_{\rm
sca})$\ the scattering form factor \citep{ML91}.  For homogeneous
spherical grains, the form factor is given by
$\Phi^2(\theta_{\rm sca}) = 3(\sin u - u \cos u)/u^3$
where $u = 4\pi a \sin(\theta_{\rm sca}/2)/\lambda \approx 2\pi a
\theta_{\rm sca} E/hc$.  

\citet{SD98} showed that the RG approximation will overestimate the
total scattering if the energy of the X-rays (in keV) is not
substantially larger than the size of the dust grains (in $\mu$m), and
suggested 2 keV as a minimum energy for most ISM dust models.

By integrating the scattering cross section over the line of sight
geometry, the source spectrum, and the dust size distribution we get
(considering single scatterings only) the halo surface brightness at
angle $\theta_h$\ from the source:
\begin{eqnarray}
I_{\rm sca}(\theta_h) & =& F_X N_H \int dE\,S(E) \int da\,n(a) \times
\\ \nonumber 
 & & \int_0^1 dx\,{{f(x)}\over{(1-x)^2}}
   {{d\sigma(E,a,\theta_{sca})}\over{d\Omega}} 
\end{eqnarray}
where $F_X$\ is the total source flux, \NH\ is the hydrogen column
density, $S(E)$\ is the (normalized) X-ray spectrum, $n(a)da$\ is the
dust grain size distribution, and $\theta_{sca} \approx
theta_h/(1-x)$.  Here $f(x)$\ is the density of hydrogen at distance
$xD$\ from the observer divided by the line of sight average density,
where $D$\ is the distance to the source \citep{ML91}.

\subsection{Multiple Scattering\label{subsec:multiscat}}

If the column density is sufficiently large, individual X-rays may be
scattered multiple times.  \citet{ML91} showed that for $\tau_{\rm
sca} > 1.3$, there are more multiply scattered photons than singly
scattered.  Multiple scattering tends to broaden the halo in an
energy-dependent fashion.  The scattering cross section itself depends
upon the X-ray energy and the dust model; Table 1 of \citet{ML91}
shows that $\sigma_{\rm sca} = 9.03\times10^{-23} E^{-2}_{\rm keV}$
for dust models such as \citet[][MRN77]{MRN77} or \citet{DL84}.  For
\NH $\approx 4\times10^{22}$\,cm$^{-2}$\ (see \S\ref{subsec:CO21}),
this corresponds to $\tau_{\rm sca} = 3.6 E_{\rm keV}^{-2}$.  Below 2
keV, we therefore expect both that the RG approximation will
overestimate the total halo intensity and that the single scattering
approximation will underestimate the radial extent of the halo.  At 2
keV, however, the RG approximation is adequate and although multiple
scattering does not dominate, it will still observably broaden the
halo.  By 3 keV, however, single scattering entirely dominates.

Explicitly calculating all possible scatterings is difficult, but it
is possible to calculate the radial profile including both single and
double scattering.  Both \citet{ML91} and \citet{PK96} derived the
general form for the total cross section including multiple dust
scattering.  Using the Gaussian approximation to $d\sigma/d\Omega$\
derived in ML91 in order to simplify the calculations, the
singly-scattered profile is
\begin{eqnarray}
I^{(1)}(\theta, E) &= &N_H c_1 \big({{\rho}\over{3}}\big)^2
\int_{x_0}^{x_1} {{dx}\over{(1-x)^2}} \int_{a_0}^{a_1} da n(a) a^6
\times \\ \nonumber 
& & \exp\big[ -0.4575 E^2 a^2 {{\theta^2}\over{(1-x)^2}}\big]
\label{eq:RGgauss}
\end{eqnarray}
where $c_1 = 1.1$\,cm$^2$sr$^{-1}$, $\rho$\ is the grain density, in g
cm$^{-3}$, $a$ the grain size in $\mu m$, $n(a)$\ is the dust grain
size distribution, and $E$\ the X-ray energy in keV.  This Gaussian
approximation to the Rayleigh-Gans cross section is used only in this
section to estimate the effect of multiple scattering; otherwise we
use the full Rayleigh-Gans model with the form factor for spherical
grains described by Equation~\ref{eq:RG}.

Equation~\ref{eq:RGgauss} can be extended using the recursion relation
given in \citet{PK96} to doubly scattered photons.  This result holds
for smoothly distributed dust between the $x_0$\ and $x_1$.  In the
following, we take advantage of the small-angle nature of the
scattering to simplify some of the trigonometry.  As a result, the
scattering over $\theta'$\ does not include all possible values of
$\theta'$\ but is limited to $\theta_1 \approx 20'$.  With that
restriction, the profile for doubly-scattered photons is:
\begin{eqnarray}
I^{(2)}(\theta, E) & = & N_H^2 c_1^2 \big({{\rho}\over{3}}\big)^4
\int_{x_0}^{x_1} dx \int_x^{x_1} {{dx'}\over{(1-x')^2}}
\int_0^{\theta_1} d\theta' \theta' \times \\ \nonumber
 &  & \int_0^{2\pi} d\phi \int_{a_0}^{a_1} da a^6 n(a) \times \\ \nonumber 
 &  & \exp\big[-0.4575 E^2 a^2 (\theta'^2 +
{{2\theta \theta'}\over{1-x}} \sin\phi) + ({{\theta}\over{1-x}})^2
\big] \times \\ \nonumber
 & & \int_{a_0}^{a_1} da' a'^6 n(a') \times \\ \nonumber 
& & \exp[-0.4575 E^2 a'^2 (\theta' {{1-x}\over{1-x'}})^2] \exp[-N_H (x'-x) \sigma_{sca}] 
\end{eqnarray}
\begin{figure*}
\includegraphics[totalheight=2.2in]{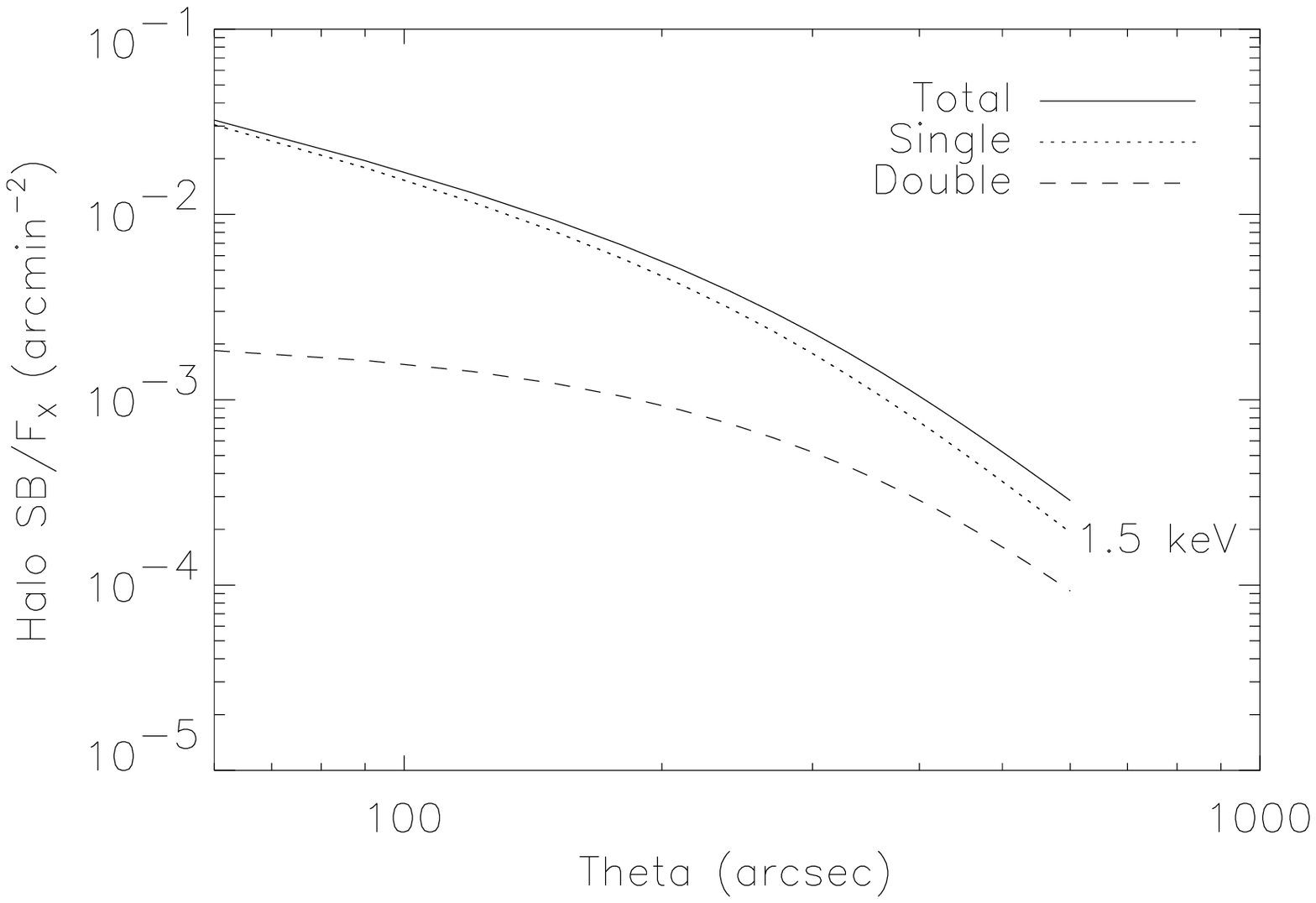}
\includegraphics[totalheight=2.2in]{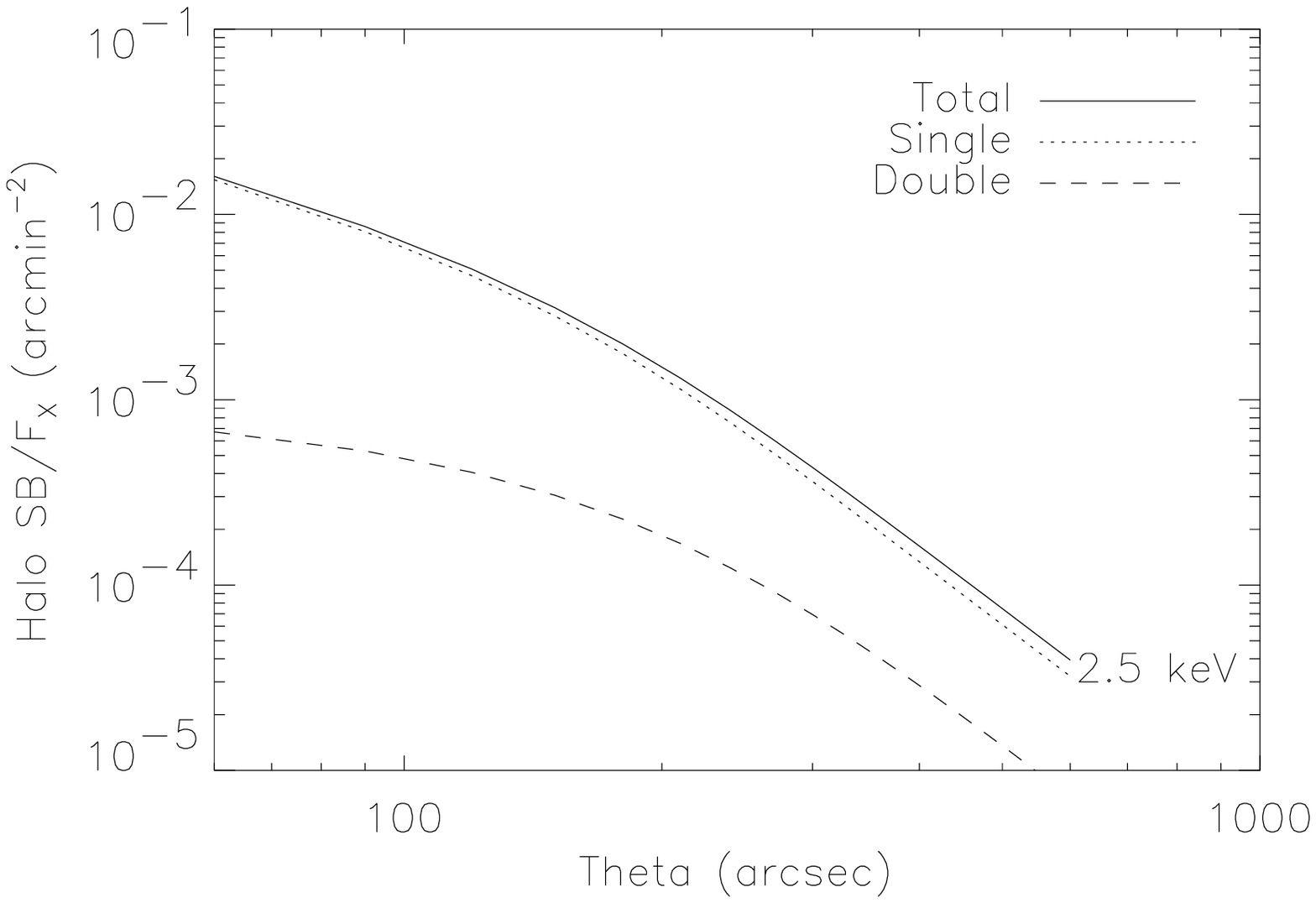}
\caption{Singly- and doubly-scattered surface brightness profiles
through a column density of N$_{\rm H} = 4\times10^{22}$\,cm$^{-2}$\
for 1.5 keV [Left] and 2.5 keV [Right] X-rays, through a smoothly
distributed MRN77 dust distribution from the observer to 90\% of the
distance to the source.\label{fig:multiscat}}
\end{figure*}
To our knowledge, this has no analytic solution.  However, it can be
evaluated numerically if $x_1 \ne 1$ (otherwise the solution becomes
numerically unstable, although the apparent singularity may only be
due to the choice of coordinates).  We considered smoothly distributed dust
between the observer ($x_0 =0$) and 90\% of the distance to the source
($x_1 = 0.9$).  The scale of the effect can be seen in
Figure~\ref{fig:multiscat}, which shows the single and
double-scattering profiles for $\NH = 4\times10^{22}$\,cm$^{-2}$\ at
1.5 and 2.5 keV, assuming an MRN77 dust model.  The total power in the
doubly-scattered photons is not insignificant.  The column density
will produce an overall optical depth for scattering $\tau = 1.25$\ at
1.5 keV and $0.42$\ at 2.5 keV.  At 1.5 keV, $\sim 50$\% of the
scattered photons will be multiply scattered, and at 2.5 keV, $\sim
20$\% \citep{ML91}.  However, the effect it has on the profile is
significant only at large radii and lower energies.  At 2.5 keV, even
at large radii, the broadening effect is relatively weak.  Due to the
twin difficulties of calculating the double scattering and the Mie
cross sections we will focus on energies $\gtrsim 2.5$\,keV.

\section{Data Reduction}

GX5-1 was observed by the \chandra\ ACIS-S for 7 ksec on August 6, 2000
(ObsID 109) and $\sim 1.5$\ million counts were detected.  Due to the
design of the \chandra\ ACIS detector, event ``pileup'' is a problem
for bright sources \citep{Davis01}.  For GX5-1, the pileup is so
severe that no counts were detected within a 2.5'' radius circle
centered at 18:01:08.217, -25:04:41.34 (J2000), after correcting the
aspect solution following the \chandra\ website
\footnote{http://cxc.harvard.edu/cal/ASPECT/fix\_offset/fix\_offset.cgi}.
We used a similar procedure to that described in \citet[][SES02]{SES02}
to determine the source position for this observation, and estimate
the error in this procedure to be $\sim$0.3'', in addition to the
0.6'' (90\% confidence) error in the \chandra\ aspect reconstruction.
Our result is 1.24'' from the position determined for the IR
counterpart (18:01:08.222, -25:04:42.58, J2000) by \citet{Jonker00}
and 0.7'' from the ATCA position (18:01:08.233, -25:04:42.044, J2000)
measured by \citet{Berendsen00}.

\subsection{Pileup}
\begin{figure}[ht]
\includegraphics[totalheight=1.6in]{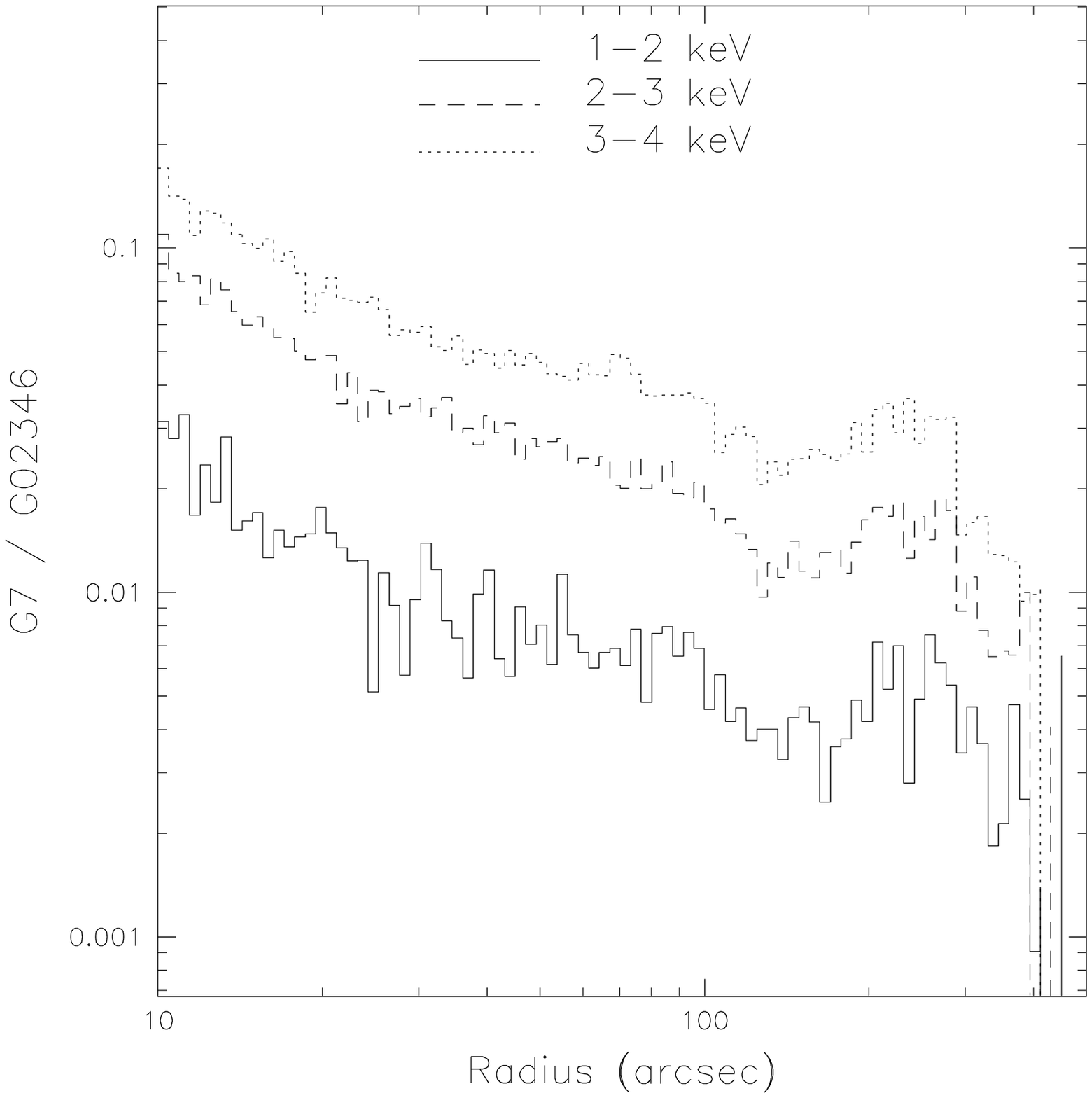}
\includegraphics[totalheight=1.6in]{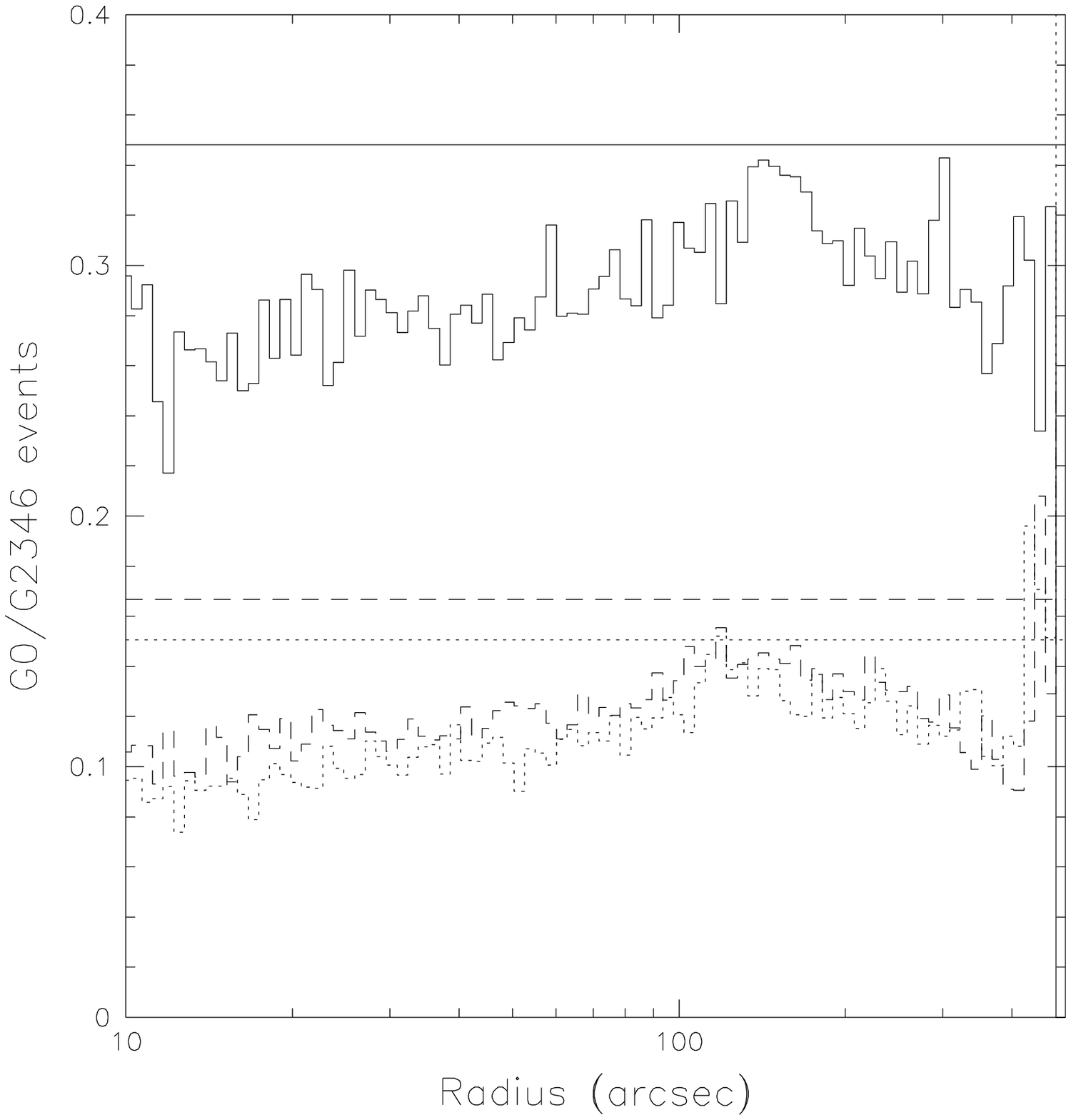}
\caption{(a) Observed radial profile of the ratio of grade 7 events to
grade 0, 2, 3, 4, and 6 events on ACIS-S3, in three bands: 1-2 keV,
2-3 keV, and 3-4 keV.  Near the source, pileup reduces the number of
good events (defined as those with grades 0, 2, 3, 4, or 6) as
multiple photons are treated as a single multi-pixel event. (b) Same,
for grade 0 to grades 2, 3, 4, and 6.  For comparison purposes the
average ratio measured in Cas A is shown as a flat
line. \label{fig:pileup}}
\end{figure}
We now have to measure the radial extent of the pileup, to determine
where we can begin to extract radial profiles with confidence.  One
way to measure this effect is to examine the ratio of 'unphysical'
grade 7 (which have detected charge in five or more pixels out of a
3x3 region) to the normally allowed grades 0,2,3,4 and 6 (which have
charge in only one to four pixels) as a function of energy and angular
distance from the source.  As the pileup diminishes, the ratio should
approach a constant value if the counts are still dominated by the
source, and then transition to the 'background' value far from the
source, as shown in Figure~\ref{fig:pileup}(a) for three energy bands.
In Figure~\ref{fig:pileup}(b) we show the ratio of grade 0
(single-pixel events) to grades 2,3,4, and 6 for the same three energy
bands.  In both cases, the ratios seem to indicate that within 100''
of the source, pileup is affecting the grade ratios.  Beyond $\sim
250''$\ the ratio might change, possibly due to the increasing
fraction of background photons.  We limited our analysis to radii
beyond $100''$\ to avoid possible contamination.

\subsection{Spectrum\label{subsec:spec}}

Measuring the spectrum of a strongly piled-up source is difficult,
since the source counts cannot be directly extracted.  The so-called
``transfer streak,'' created by mispositioned counts that arrived
during the frame transfer provide the only clean unpiled spectrum in
the data.  The transfer streak has not yet been explicitly calibrated
by the \chandra\ X-ray Center team.  Nonetheless an accurate spectral
measurement is crucial when subtracting the instrumental PSF as well
as calculating the total dust column density. Our approach is similar
to that described in SES02, except for the treatment of the
background.  The source counts were extracted from a 8 pixel wide by
611 pixel long box that stretched across the CCD, with the exception
of the region within 100'' of GX5-1 (203 pixels), which was excluded
to avoid pileup.  The allowed region contained 48465 counts, with a
total observing time of 6868.7 seconds, or a total of 2146 3.2 second
CCD ``frames.''  Each CCD frame, therefore, had $\sim 22.5$\,counts in
the transfer streak, spread over a 8$\times$611 pixel box so pileup in
the readout streak is not significant.  Since the 8 pixel wide box
contains essentially all of the transfer streak, the total effective
exposure time for the transfer streak is 2146 frames $\times$\ 611
rows/frame $\times 41\,\mu$s/row = 53.76 seconds.

It is important to note that this is the exposure time for the {\sl
source}\ counts.  The background flux (measured using boxes parallel
but above and below the source boxes) is caused by scattered photons
from GX5-1, cosmic rays, and other sources.  These events could arrive
during the transfer streak, but are much more likely to occur during
the 3.2 s hold time, so the appropriate exposure time is the full
6868.7 seconds.  This is an unusual situation: the exposure times for
the source and background photons in the source spectrum are
different, and none of the standard spectral tools (XSPEC or Sherpa)
can explicitly deal with it.  We dealt with this problem by treating
both the source and background as having the full exposure time of
6868.7 seconds, and then scaling the source flux by the ratio of the
exposure times, $6868.7/53.7495 = 127.767$.

After extracting the source and background spectra, we found an
adequate fit using an absorbed bremsstrahlung model ($\chi^2_{\nu} =
1.2$) with $\NH = 4.0\times10^{22}$\,cm$^{-2}$, $kT = 10.5$\,keV, and
$F_X$(1-10 keV)$ = (5.2\pm0.1)\times10^{-8}$\,erg/cm$^2$/s (absorption
corrected).  Fortuitously, there was also a 9 ksec HETG observation of
GX5-1 (Obsid 716) done July 18, 2000, only 19 days before the ACIS
observation described here.  Although GX5-1 is a variable source, the
XTE All-Sky monitor shows little change between the two observations.
In the 1.3-3.0 keV channel, the RXTE count rate was $14.2 \pm 1.05$\
during the HETG observation and $13.4 \pm 1.01$\ during our ACIS
observation.  In the 5.0-12.1 keV channel, the count rates were
$32.3\pm1.3$\ and $35.1\pm1.3$, respectively.

The HETG observations of GX5-1 have been analyzed by \citet{Ueda05} in
order to measure the elemental absorption along the line of sight.  As
noted in \S 1, they found the total \NH $=
2.8^{+3.3}_{-1.8}\times10^{22}$\,cm$^{-2}$, based on extrapolation
primarily from Mg, Si, S, and Fe absorption features.  Their spectral
model was more complicated than ours, but their total (absorption
corrected) $F_X$(1-10 keV)$ = 4.3\times10^{-8}$\,erg/cm$^2$/s.  This
is $\sim 20$\% lower than our result from the transfer streak which
may be due to true variation in the source.  However, given
the limited data available for the ACIS transfer streak, this 20\%
variation may also be due to calibration uncertainties.  In any event,
although this overall flux uncertainty will be reflected in the
measured dust column density, but does not affect the profile of the
halo.

\subsection{Imaging Analysis}

To calculate the radial profile of GX5-1, we first identified the
serendipitous sources from the image, using the CIAO routine {\tt
celldetect}.  We then selected an energy grid (1-4 keV, $\Delta
E=0.2$\,keV) and a radial grid (100 points from 10-800'', log-spaced).
We filtered the data using the measured CCD energy, and extracted the
counts in the radial profile using concentric annuli centered on the
source.  After making exposure maps for each energy band, we filtered
out the regions with serendipitous sources in both the data and each
exposure map.  We then extracted the effective ``radial exposure''
using the same concentric annuli, for each energy band.  Dividing the
radial profile of the counts in each band by the effective area and
exposure time gave the corrected radial profile in units of photons
cm$^{-2}$ s$^{-1}$ arcmin$^{-2}$\ per energy band.  We then divided
this by the observed source flux in each band to get the fractional
radial profile in units of source fraction arcmin$^{-2}$.  We fit this
as the sum of the scattered halo, the \chandra\ PSF, and the background
(both instrumental and cosmic).  As with SES02, the \chandra\ PSF was
measured from observations of Her X-1.  We also note that the \chandra\
PSF is small enough that we can neglect scattering of the X-ray halo
by the telescope.  Finally, although the cosmic background is
vignetted while the instrumental background is not, we found that in
these bandpasses we could simply fit them together as a flat continuum
out to $800''$.

\subsection{CO and 21 cm observations\label{subsec:CO21}}
\begin{figure}[t]
\includegraphics[totalheight=2.3in]{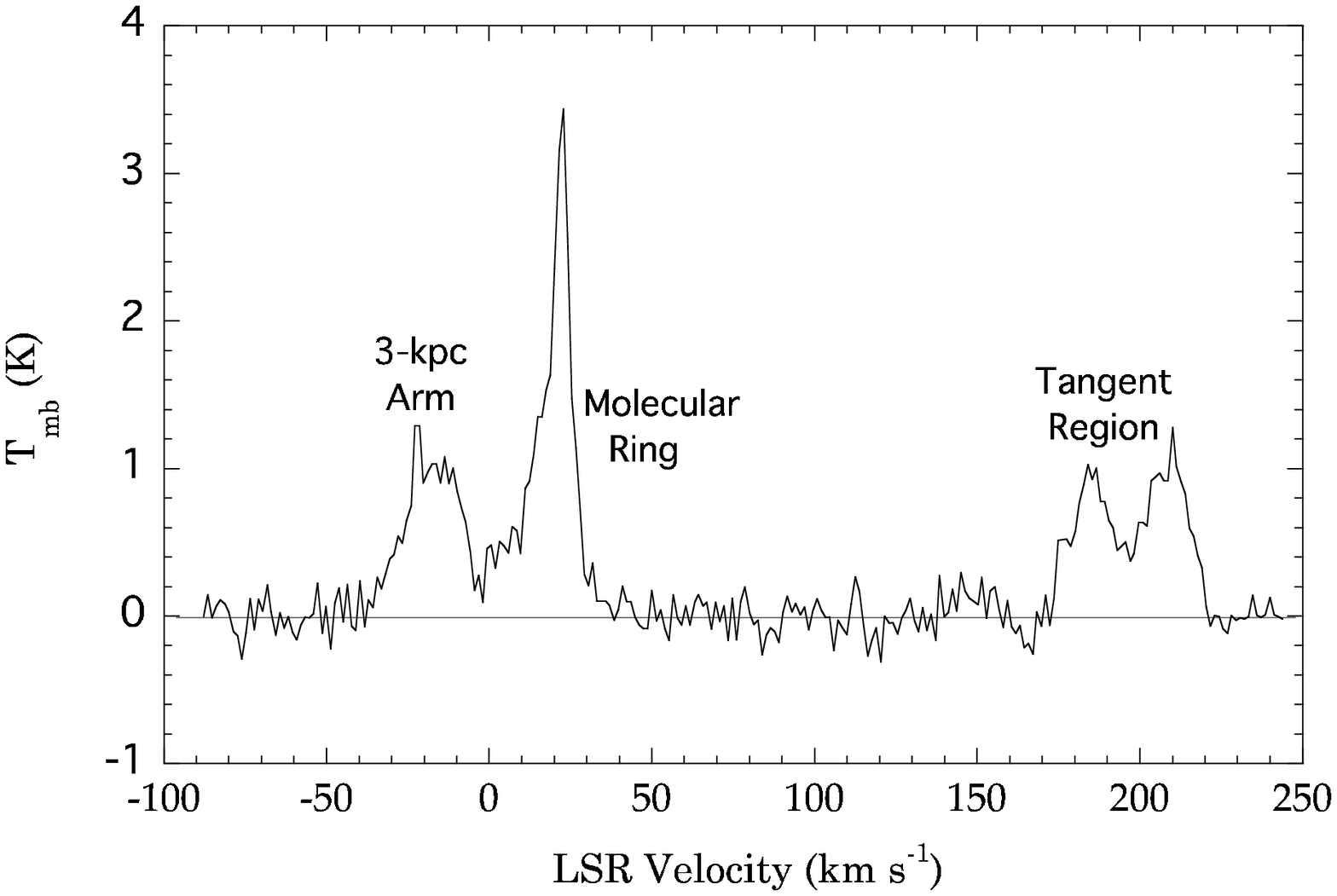}
\caption{The CO emission observed in the direction of
GX5-1 \citep{Dame01}.\label{fig:CO}}
\end{figure}
\begin{table*}
\caption{Molecular Concentrations toward GX5-1 from CO
observations\label{tab:cloud}}
\begin{center}
\begin{tabular}{llllll}
\hline \hline 
Cloud&Velocity&W$_{\hbox{CO}}$&2$\times\hbox{N}(\hbox{H}_2)$ &N(H{\sc
i}) &Distance from Sun \\ 
     &  (km/s)& (K km/s) & ($10^{20}$\,cm$^2$)& ($10^{20}$\,cm$^2$) &
(kpc) \\ \hline 
Molecular Ring&  22    &   40.4   &    146    & 34 & 4.7 (or 3.3; see
\S\protect{\ref{subsec:CO21}}) \\ 
3 kpc Arm     & -22    &   24.4   &     88 (or 18$^{\dag}$) & 34 & 5.1  \\ 
Galactic Center (Tangent) region &184, 208&   21.7   &16$^{\dag}$& 11 & 8.5 \\ 
Total Line of Sight &  &   86.5   &    250 (or 180) & 83 &      \\ \hline \hline
\end{tabular}
\end{center}
$^{\dag}$We have reduced the expected H$_2$\ by a factor of 5, since
molecular gas near the Galactic Center is thought to be anomalously
bright in CO\citep[{e.g.}][]{Sodroski95}.
\end{table*}
The CO emission toward GX5-1 (Figure~\ref{fig:CO}) is confined to 4
broad lines that can be readily associated with three distinct regions
of the Galaxy. First, the line peaking near -22 km~s$^{-1}$\ almost
certainly arises from the 3 kpc Expanding Arm \citep[e.g.][]{Bania80},
one of the best-defined spiral arms in the Galaxy.  It is seen
tangentially at $l \sim 24^{\circ}$, implying a Galactic radius of
3.46 kpc (assuming a distance to the Galactic Center of 8.5
kpc). Since the inclination of the arm is unknown, we assumed it is
circular to calculate its distance as 5.1 kpc at $5^{\circ}$\
longitude.  A second region is associated with the line at 22
km~s$^{-1}$, which arises from the so-called molecular ring.  This is
the region of high molecular cloud density roughly half-way between
the Sun and the Galactic center. The near and far kinematic distances
of this gas are 4.7 kpc and 12.1 kpc, with uncertainties of
$\pm0.5$\,kpc imposed by the cloud-cloud velocity dispersion. At the
far distance the molecular gas would lie 215 pc above the plane, more
than 4 times the vertical dispersion of molecular clouds
\citep{Bronfman88}, so the near distance is far more likely.

It is worth noting that most of the molecular ring emission closer to
the plane at $l\sim 5^{\circ}$\ is centered roughly 10 km~s$^{-1}$\
lower than that toward GX5-1; this (mildly) suggests that the
molecular-ring emission toward GX5-1 might be anomalously high by that
amount. If so, its kinematic distance should be reduced to $\sim
3.3$\,kpc; this possibility is also noted in Table~\ref{tab:cloud}.
We conclude that the molecular ring gas probably lies 3-5 kpc from the
Sun. Given the extreme velocity crowding at this low longitude, it's
possible that numerous clouds are spread over this range.  A third
region is associated with the other two partially-blended lines. These
lines are at such high velocities ($\sim$\ 200 km~s$^{-1}$) that the
emitting region must lie near the tangent point, at a Galactic radius
of 0.7 kpc and thus near the Galactic center at a distance of $\sim
8.5$\,kpc. Distances for the main molecular concentrations along the
line of sight to GX5-1 are summarized in Table~\ref{tab:cloud}.

The molecular column density, N(H$_2$), corresponding to each of the
CO lines can be calculated from their velocity-integrated intensities,
W$_{\rm CO}$.  For the molecular ring emission we use a standard
Galactic value for $X \equiv$\ N(H$_2$)/W$_{\rm CO} =
1.8\times10^{20}$\,cm$^{-2}$K$^{-1}$km$^{-1}$s \citep{Dame01}.  Since
there is evidence from both diffuse gamma ray emission \citep{Blitz85}
and far infrared dust emission \citep{Sodroski95} that molecular
clouds in the Galactic center region are anomalously bright in CO, we
reduced $X$\ by a factor of 5 for the two high-velocity lines near the
Galactic center.  The molecular cloud in the 3 kpc Arm might also be
anomalously bright in CO, given the arm's proximity to the Galactic
center and large non-circular motion.  Our results are given in
Table~\ref{tab:cloud}, along with the corresponding atomic column
densities derived from the 21 cm survey of \citet{Hartmann97}. The
interpolated 21 cm spectrum toward the source shows velocity
components similar to those seen in CO. We found a total \ion{H}{1}
column density along the line of sight of
$0.83\times10^{22}$\,cm$^{-2}$, in good agreement with the
\citet{DL90} value of $0.91\times10^{22}$\,cm$^{-2}$. The total gas
column density, $2\times$N(H$_2$) + N(\ion{H}{1}), is
$3.3\times10^{22}$\,cm$^{-2}$, or $2.6\times10^{22}$\,cm$^{-2}$,
depending upon the value of $X$\ used for the 3 kpc Arm gas (see
Table~\ref{tab:cloud}).

As with many X-ray binaries, the true distance to GX5-1 is not well
determined.  We use the \citet{Christian97} upper limit of 9 kpc,
based on its flux and calculated Eddington luminosity.  Comparing this
upper limit with Table~\ref{tab:cloud}\ shows that GX5-1 is almost
certainly behind the 3 kpc arm and the molecular ring, but may be in
front of, behind, or embedded in the Galactic center region.  The
radial profile of the X-ray halo depends on the dust position relative
to the source flux, but since X-ray halos are due to forward
scattering, dust behind the source is unimportant.

\section{Results and Discussion}

\begin{figure*}
\includegraphics[totalheight=2.3in]{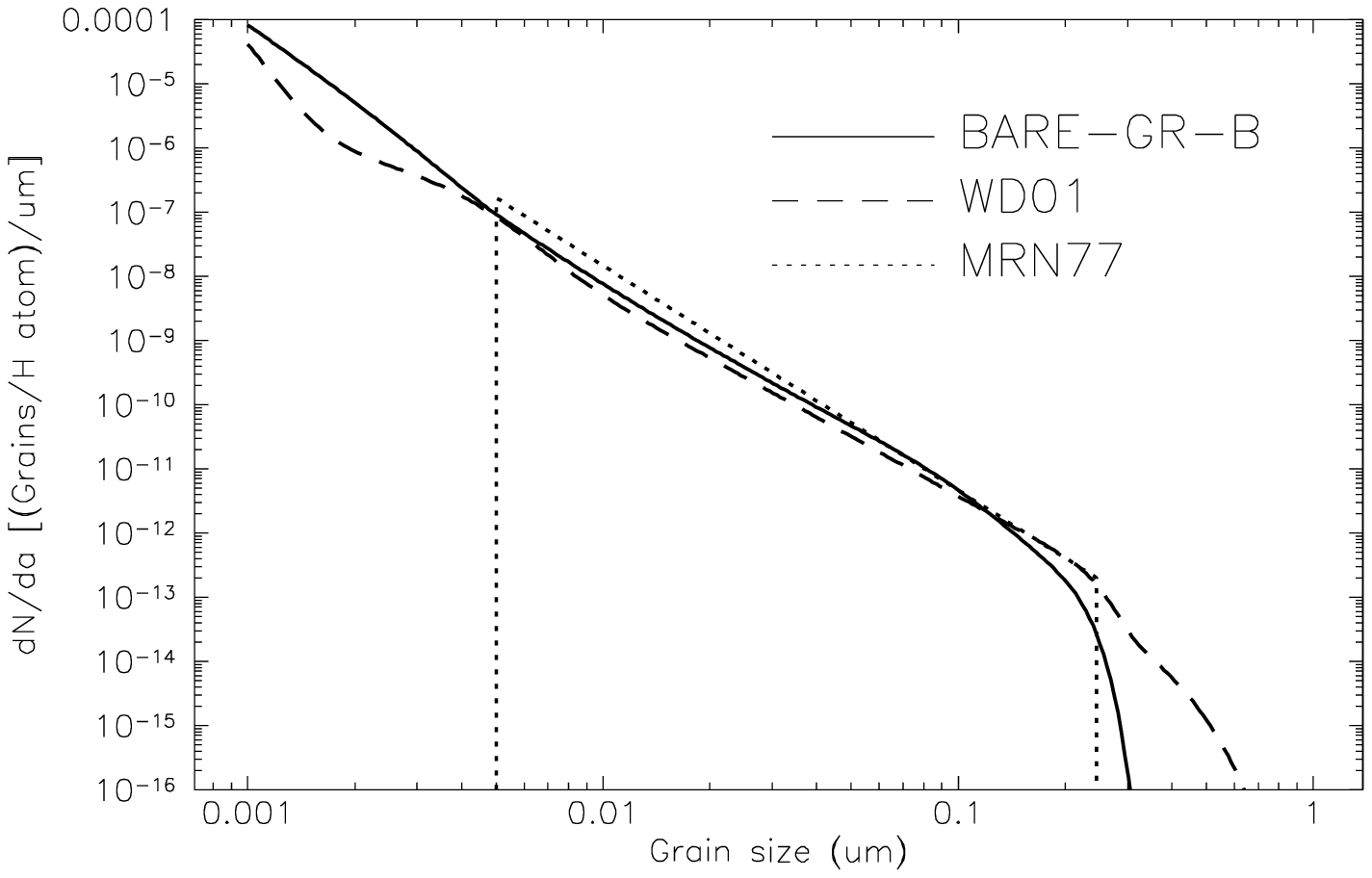}
\includegraphics[totalheight=2.3in]{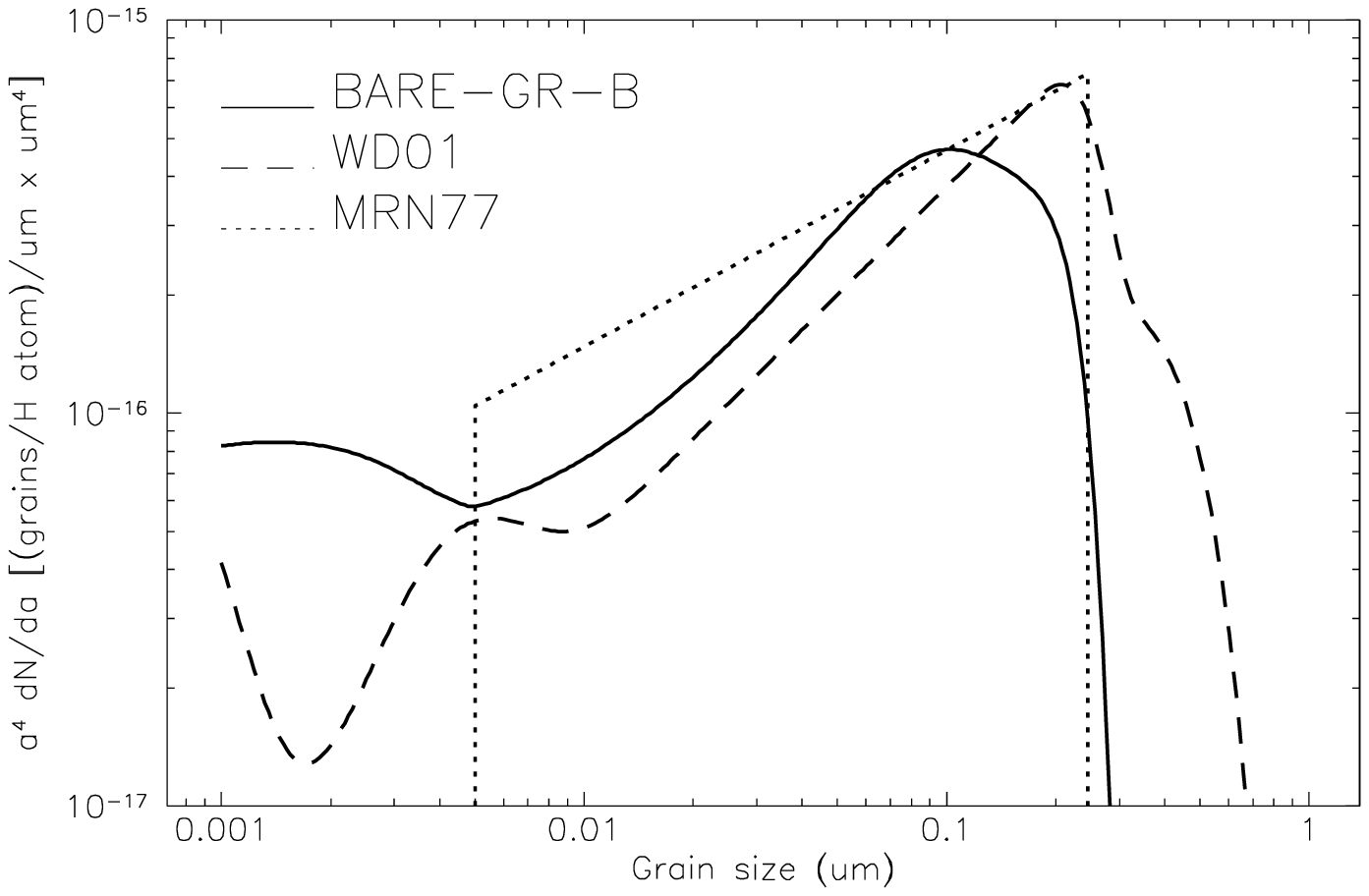}
\caption{[Left] Total dust grain size distributions (summed over all
  components) for the ZDA04 BARE-GR-B model (solid line), the WD01
  R$_V=3.1$\ model (dashed line), and the MRN77 model (dotted line).
  [Right] Same, but weighted by the RG X-ray total scattering cross
  section factor $a^4$. \label{fig:SizeDist}}
\end{figure*}

GX5-1 is one of the brightest persistent X-ray sources in the Galaxy,
and its relatively large absorption column density means that it has a
substantial X-ray halo.  The CO and \ion{H}{1} observations also
constrain the fit parameters.  As a result, GX5-1 is one of the best
sources to use in testing IS dust models.  We therefore collected a
wide range of proposed dust models: MRN77, \citet[][WD01]{WD01}, and
the 15 models proposed in \citet[][ZDA04]{ZDA04} (most notably their
BARE-GR-B model), as well as the model including extremely large
grains discussed in \citet[][WSD01]{WSD01}.  These models were
developed to match optical, UV, and IR observations of dust primarily
found in diffuse clouds (or, in the case of WSD01, {\it in-situ}\
data) without consideration of their X-ray scattering properties.
Figure~\ref{fig:SizeDist}[Left] shows the size distributions for three
models, while Figure~\ref{fig:SizeDist}[Right] shows the size
distributions weighted by $a^4$, making them proportional to the total
X-ray scattering cross section following Equation~\ref{eq:RG} (after
integrating over all scattering angles; see \citet{ML91}).  This shows
that the differences in the largest grains must dominate the results.
By systematically comparing all of these to the GX5-1 data, we can
determine which models agree with the observed X-ray scattering, and
which do not.  However, it is important to realize that the dust
grains scattering the X-rays from GX5-1 are primarily in dense clouds,
and may have quite different properties from the dust in diffuse
clouds.  Therefore, this study cannot globally exclude dust models,
but simply determine which could describe the dust in dense clouds and
which cannot.
Our results are additionally limited by the extreme pile-up, which
obscures the true profile when $\theta < 100''$.  Dust very near the
source ({\it e.g.}\ the clouds in the Galactic center region) will
lead to scattered photons in our insensitive ``near region'' (see also
Figure~\ref{fig:NearProfile}) In addition, the number of scattered
photons at every angle is proportional to $F_X N_H$, so errors in the
measured flux linearly correlate with errors in the dust column
density.  Finally, the total column density towards the source is not
well-known.  The measurement by \citet{Christian97} of \NH $= 2.54 \pm
0.19 \times 10^{22}$\,cm$^{-2}$\ is reasonably precise, but its
accuracy depends entirely on the spectral model assumed.  The same is
true of our best-fit value \NH $=4.0\times10^{22}$\,cm$^{-2}$ (see
\S\ref{subsec:spec}), with the additional caveat that calibrating a
transfer streak spectrum is challenging.  The \citet{Ueda05} result of
\NH $= 2.8^{+3.3}_{-1.8}\times10^{22}$\,cm$^{-2}$, based on absorption
features seen in high-resolution spectra, is accurate but far less
precise.  Considering the uncertainties on the H{\sc i}\ large-scale
optical depth the CO-to-H$_2$\ conversion factor, the value of $\NH =
3.3\times10^{22}$\,cm$^{-2}$\ derived from the gas tracers is probably
only good to within a factor of 2.  Based on all these results, the
most we can say with confidence is that \NH\ is likely in the range
$2-4\times10^{22}$\,cm$^{-2}$.  If our X-ray halo results are outside
this range, it would imply is that the dust/gas mass ratio along this
line of sight is abnormally high or low.  Therefore, we cannot impart
too much significance to the overall normalization of our results.
However, we can require that the best-fit column density be
independent of energy, and that the shape of the profile conform to
the observations.

\subsection{Smoothly Distributed Dust}
\begin{figure*}
\includegraphics[totalheight=2.3in]{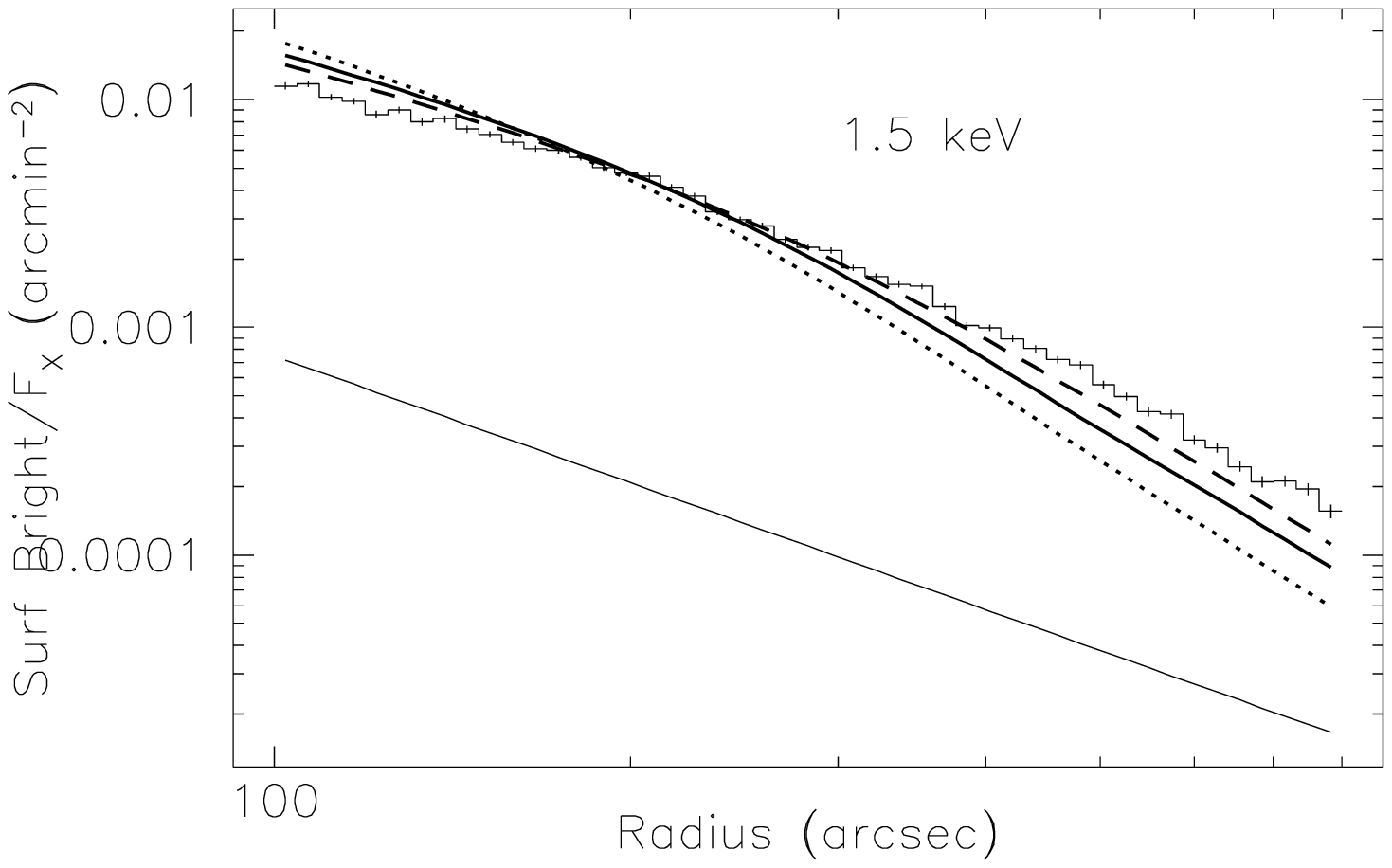}
\includegraphics[totalheight=2.3in]{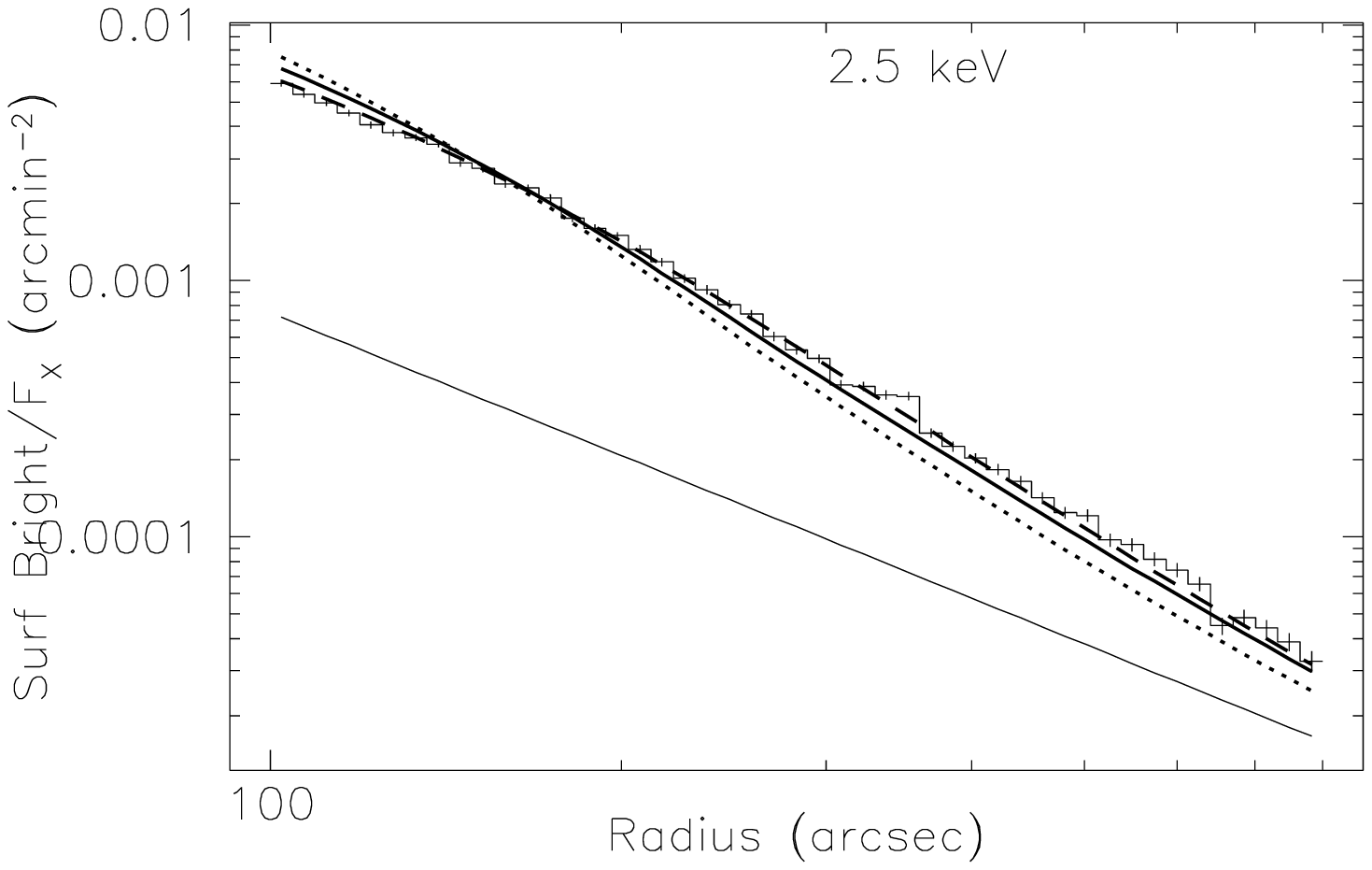}
\caption{[Left] Fractional halo profile of GX5-1 at 1.5 keV, fit with
a (single-scattering) smooth distribution using the MRN77 (solid
line), WD01 (dotted line), and ZDA04 (dashed line) dust models.
Multiple scattering is not included; including it would increase the
model scattering at large angles and decrease it at small angles. The
power-law fit to the \chandra\ PSF is also shown as a solid line.
[Right] Same, at 2.5 keV, except that multiple scattering is much less
significant. \label{fig:SmooDust}}
\end{figure*}
Although the CO data strongly implies the dust along the line of sight
is clumped, we begin with the simplest X-ray halo model, where the
dust is assumed to be smoothly distributed along the line of sight
($f(x) \equiv 1$), and that it scatters the X-rays at most once
\citep{PS95}.  We begin by showing the radial profiles fit with some
of the most common or best-fitting dust models, to show the quality of
the results.  Figure~\ref{fig:SmooDust}\ shows the radial profile of
GX5-1 at $1.5\pm0.1$\ and $2.5\pm0.1$\,keV, fit using a
single-scattering smooth dust distribution.  We show three different
dust models: the classic MRN77 model as well as the more recent WD01
(with $R_V = 3.1, b_C = 6\times10^{-5}$) and the BARE-GR-B model from
ZDA04.  The best-fit column densities were (respectively) $2.9$,
$2.1$, and $4.2\times10^{22}$\,cm$^{-2}$\,at 1.5 keV and $2.7, 2.2$,
and $3.5\times10^{22}$\,cm$^{-2}$\ at 2.5 keV.  At 1.5 keV all the
fits were inadequate, although the ZDA04 model was the best fit.
Although at 1.5 keV Mie effects could affect the halo slightly, a fit
including these did not significantly change the results.  However, as
noted in \S\ref{subsec:multiscat}, comparing with
Figure~\ref{fig:multiscat} shows that including multiple scattering
might have improved the ZDA04 model fit.  Unfortunately, including
multiple scattering from smoothly distributed dust in these fits is
not yet implemented.  However, in the case of GX5-1 the dust is almost
certainly {\it not}\ smoothly distributed and so this enhancement is
left to a future paper.  At 2.5 keV, Figure~\ref{fig:multiscat}[Right]
showed that multiple scattering is far less important, and we see in
Figure~\ref{fig:SmooDust}[Right] that the ZDA04 model fit the radial
profile of GX5-1 reasonably well-- $\chi_{\nu}^2($ZDA04$) =
1.8$.  The MRN77 and WD01 models were still poor fits with
$\chi_{\nu}^2($MRN77$) = 9.4$\ and $\chi_{\nu}^2($WD01$) = 30$.  They
both overestimated the scattering at small angles and underestimated
it at large angles far more than the ZDA04 model.
\begin{figure}
\includegraphics[totalheight=2.3in]{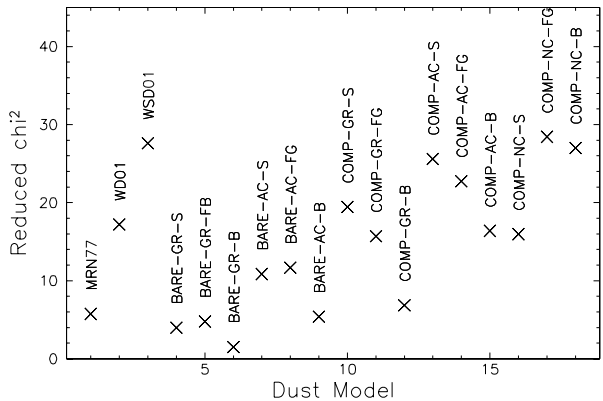}
\caption{$\chi_{\nu}^2$\ values for each dust model used, assuming
  smoothly-distributed dust along the line of sight and using the data
  between 2.5-3.5 keV in 0.2 keV bins.  At these energies, multiple
  scattering should be largely insignificant. \label{fig:SmoothChiSq}}
\end{figure}
We then fit the GX5-1 radial profile data for energies between 2.5-3.5
keV (in 0.2 keV wide bins) to a model consisting of the \chandra\ PSF,
a flat background, and the scattered halo calculated for
smoothly-distributed dust following each of the above dust models.  In
Figure~\ref{fig:SmoothChiSq}\ we show the $\chi^2_{\nu}$\ values for
each best-fit model.  While none of the models are formally
acceptable, clearly some are preferred.  For example, none of the
composite models of ZDA04 (those beginning with COMP-), with the
possible exception of COMP-GR-B, are even close to a reduced $\chi^2_{\nu}
\sim 1$, while the BARE-GR-B model is a reasonably good fit ($\chi^2_{\nu} =
1.5$), as would be expected from the results in
Figure~\ref{fig:SmooDust}.

\begin{figure*}
\includegraphics[totalheight=2.3in]{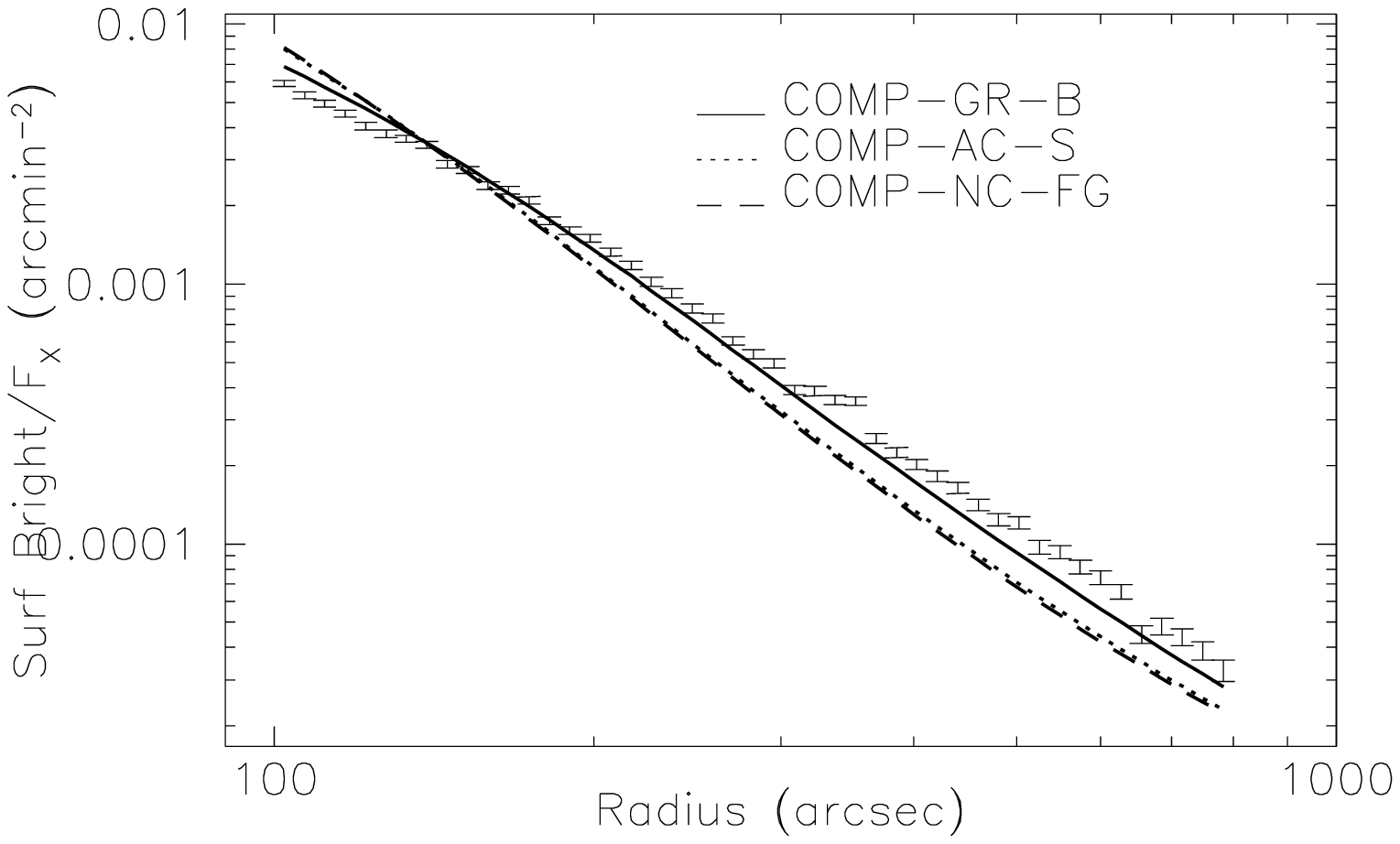}
\includegraphics[totalheight=2.3in]{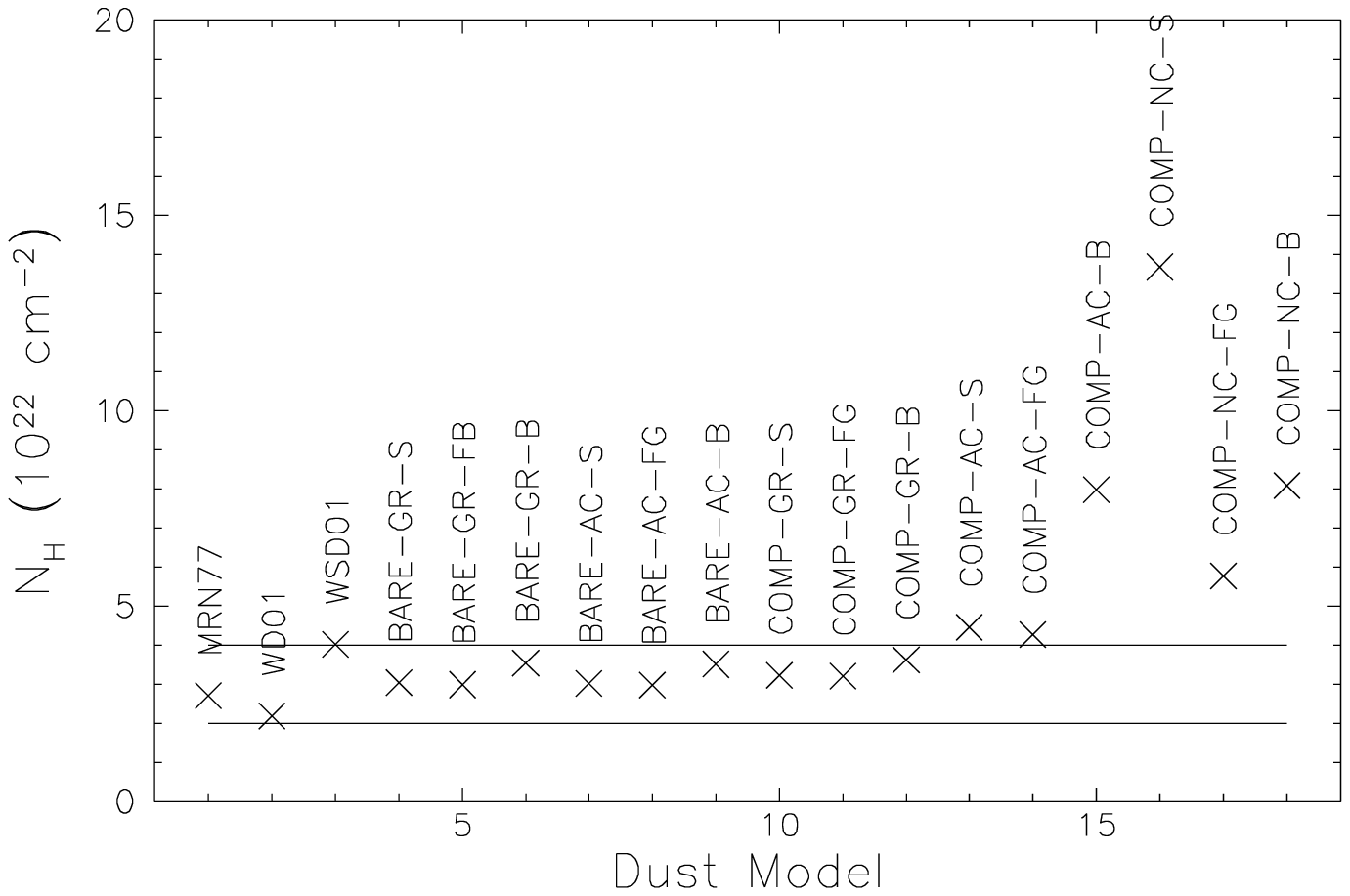}
\caption{[Left] Radial profile of GX5-1 at 2.5 keV, with the best-fit ZDA04
  COMP-GR-B, COMP-AC-S, and COMP-NC-FG models assuming
  smoothly-distributed dust.  At large scattering
  angles these models all under-predict the total scattering, while at
  smaller angles ($<150''$) the models overestimate the observed
  halo. [Right] The best-fit column density N$_H$\ for each model; the
  horizontal lines mark the expected upper and lower values.  Most
  models find a reasonable total dust model, despite an overall poor fit.
   \label{fig:SmCompProf}}
\end{figure*}

While instructive, Figure~\ref{fig:SmoothChiSq}\ is not conclusive.
First, GX5-1's brightness means that the errors are probably not
statistical but are instead dominated by calibration errors in the PSF
and effective area.  Second, the assumption of smoothly distributed
dust disagrees strongly with the CO observations and is almost
certainly incorrect.  At the same time, the consistently poor fits
found when using the composite models of ZDA04 suggests that these
models poorly describe the observations.  In
Figure~\ref{fig:SmCompProf}[Left] we show the fits to the radial
profiles using the COMP-GR-B, COMP-AC-S, and COMP-NC-FG models.  All
of these over-predict the scattering at small angles and under-predict
it at large angles. Since larger grains create smaller halos, this
suggests that these models have relatively too many large grains.
Interestingly, however, the best-fit column densities (shown in
Figure~\ref{fig:SmCompProf}[Right]) are almost all in the
$2-4\times10^{22}$\,cm$^{-2}$\ range, with the exception of the last
six models.  In general, then, these models have problems not with the
total mass of dust required, but rather the relative distribution of
large and small grains.
\subsection{Clumpy Cloud Models\label{subsec:single}}
\begin{figure}
\includegraphics[totalheight=2.3in]{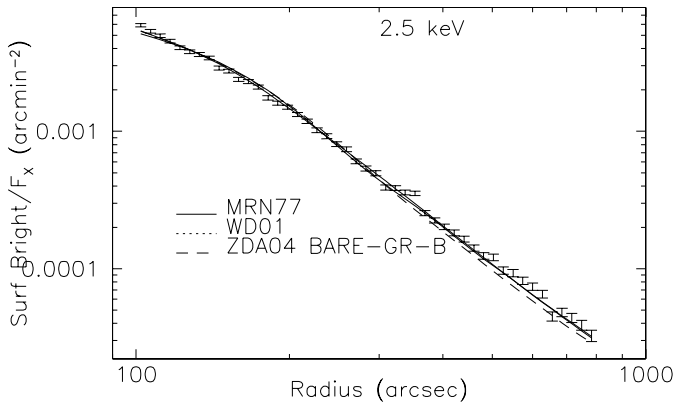}
\caption{Best-fit scattered X-ray halo to the GX5-1 data for a single
  cloud at 2.5 keV.  With the ZDA04 BARE-GR-B model, \NH $=
  2.5\times10^{22}$\ and the cloud position is 33\% of the distance to
  the source; $\chi^2_{\nu} = 2.1$.  The MRN77 and WD01 models, despite
  their similar appearance, are worse fits with $\chi^2_{\nu} = 3.7, 5.7$\
  respectively.
  \label{fig:OneCloud}}
\end{figure}
Although the good smoothly-distributed ZDA04 BARE-GR-B model fit could
be taken as evidence for that model over the others considered, we
must also consider the possibility that the match is at least
partially fortuitous as the dust is almost certainly clumped into the
same clouds seen in CO emission.  In this case the halo profile depends
upon the relative positions of the clouds and the source.  We start by
considering the simplest possible model, a single cloud with variable
relative position and column density.  The parameters were allowed to
vary freely without any assumptions taken from the CO observations.
This model again includes only single scattering.  We begin with fits
to the same three dust models (see Figure~\ref{fig:OneCloud}), and
again find the best results from the ZDA04 BARE-GR-B model.  At 2.5
keV, all models appears to fit equally well but only the ZDA04
BARE-GR-B model has plausible fit parameters.  The best-fit column
densities for the MRN77, WD01, and BARE-GR-B models were $1.7, 1.2,
2.5\times10^{22}$\,cm$^{-2}$, respectively and the best-fit dust
positions (in units of fractions of the total distance to the source) were
$0.16, 0,$\ and 0.32.  In both cases, the MRN77 and WD01 parameters
are well outside the expected values.  In addition, the BARE-GR-B
model's best-fit column density appeared to be nearly constant with
energy, unlike the other two models, and it had the lowest overall
$\chi^2_{\nu}$\ values.  The best-fit ZDA04 BARE-GR-B model had $\NH =
2.5\times10^{22}$\,cm$^{-2}$\ and position $\sim 30$\% of the distance
to GX5-1, with slight variation between $25-35\%$\ as a function of
energy.  Nonetheless, even this fit is formally unacceptable
($\chi_{\nu}^2 > 2$) at most energies.  Note that all fits strongly
rejected the result found by \citet{Xiang05}, who found $>90$\% of the
dust to be very near the source.

We now consider a more complex spatial distribution of dust clouds,
based on the CO data.  The single cloud results foreshadow possible
consistency problems between the scattered X-rays and a ``naive''
interpretation of the CO data.  As Table~\ref{tab:cloud} shows,
multiple interpretations of the CO data itself are possible.  The 3
kpc Arm may have substantial or insignificant amounts of dust, and the
Molecular Ring dust may be at either 4.7 or 3.3 kpc.  We considered
all combinations of these options, and decided to use a value of
$1.8\times10^{21}$\,cm$^{-2}$\ for the 3 kpc Arm column density and a
distance of 3.3 kpc  for the Molecular Ring, as other choices led to a
poorer fit to the data for all dust models. Of
course, this may be an indication that there is a problem either in
the calibration or the dust models, but we first examine if the closer
distance can lead to a good result. 

\begin{figure*}
\includegraphics[totalheight=2.3in]{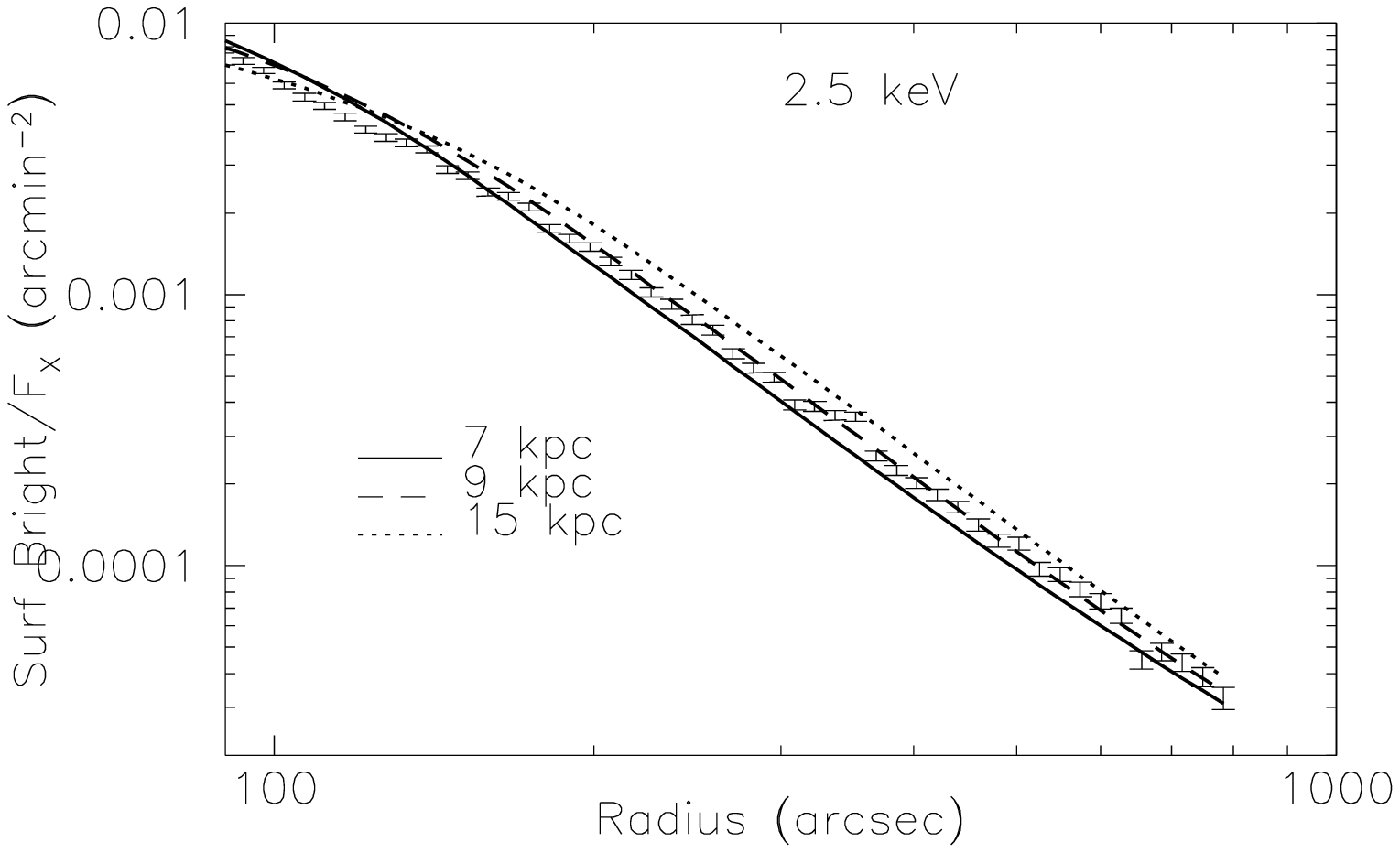}
\includegraphics[totalheight=2.3in]{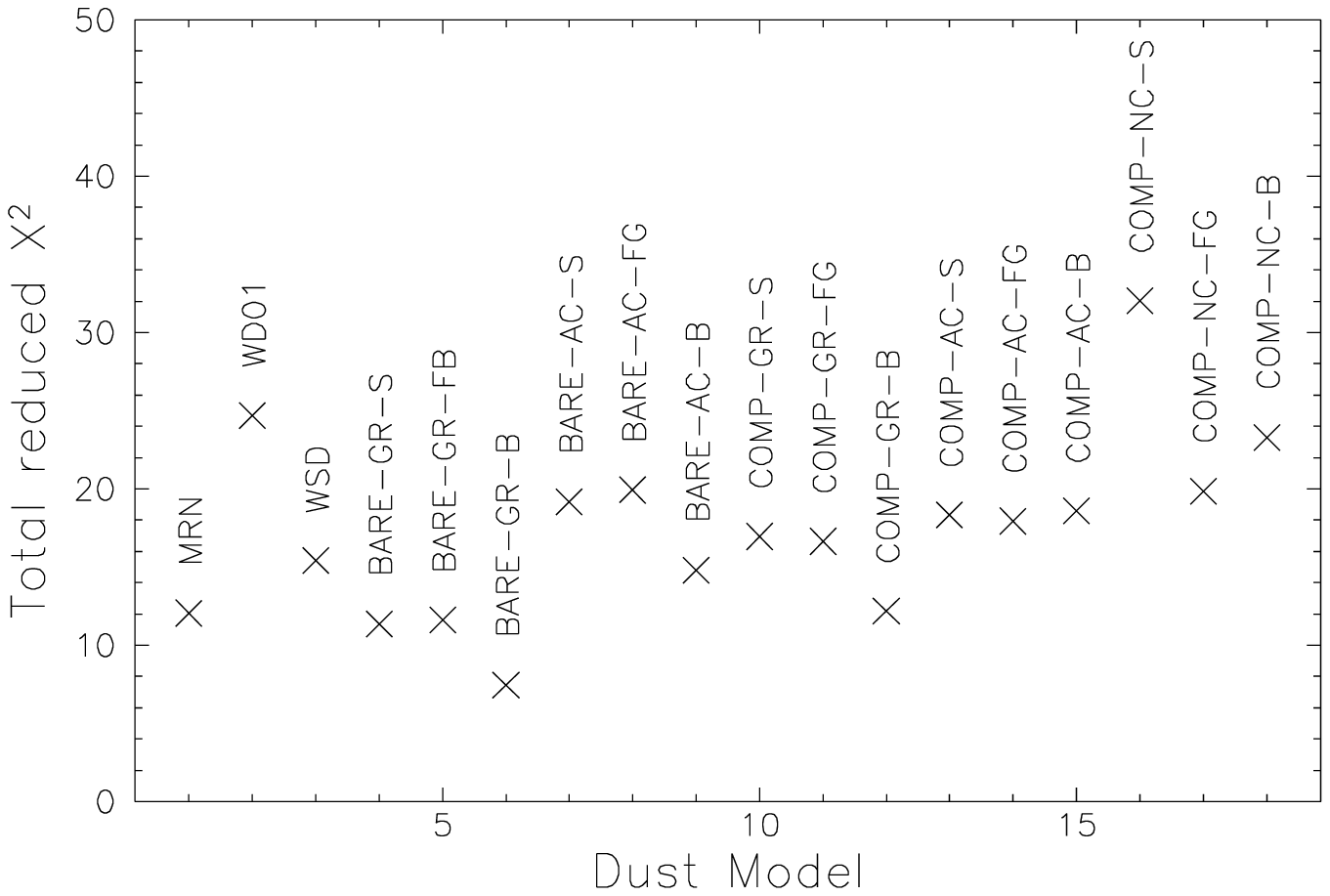}
\caption{[Left] Observed radial profile (divided by source flux) at
  2.5 keV with ZDA04 BARE-GR-B dust model using column densities and
  positions derived from the most plausible interpretation of the CO
  data with 3 possible source positions.  [Right] $\chi_{\nu}^2$\
  values for fits to the data between 2.5-3.5 keV (in 0.2 keV bins)
  for each dust model, assuming the same parameters and a source
  distance of 8 kpc.
  \label{fig:MultiCloud}}
\end{figure*}
In Figure~\ref{fig:MultiCloud}[Left] we show the halo profile fit
using these preferred values from the CO observations to fix the cloud
column densities and positions, and show the expected halos from a
ZDA04 model considering single scattering only with three different
positions.  In this case the only variable fit parameter is the source
flux.  For these fits, the best-fit value is within 25\% of the flux
measured from the transfer streak, within the calibration uncertainty.
Once again, the ZDA04 model fits were superior to either the
MRN77 or WD01 models, and so for clarity only the ZDA04 model is
shown.  A distance of 15 kpc is clearly a poor fit, over-predicting the
halo brightness at nearly all angles (a distance of $\sim 15$\,kpc for
GX5-1 would also imply the source has a consistently super-Eddington
luminosity \citep{Christian97}).  The best-fit appears to be bracketed
between the 7-9 kpc distances, although neither of these is truly an
acceptable fit.  Only single-scattering was considered for these
plots; multiple scattering would tend to broaden the halo slightly and
might improve the fit slightly for the 7 kpc distance.  However, as
shown in Figure~\ref{fig:multiscat}[Right], the improvement would only
be marginal and is unlikely to explain the entire difference.

In Figure~\ref{fig:MultiCloud}[Right] we again show the $\chi_{\nu}^2$\
values for each best-fit model, this time for the multiple cloud model
as suggested by the CO data.  We assumed a distance for GX5-1 of 8
kpc, which gave the best overall results.  As with
Figure~\ref{fig:SmoothChiSq}, none of the models are formally
acceptable, although some of the same models are again preferred, most
notably the ZDA04 BARE-GR-B model.  As with the smoothly-distributed
dust model, the composite dust models of ZDA04, especially the ``COMP-NC''
models which use No Carbon, give very poor fits.

\section{Conclusions}

We have extracted the spectrum and scattered halo from GX5-1, one of
the brightest Galactic X-ray binaries.  Although the spectrum had to
be extracted from the poorly-calibrated transfer streak, we were able
to confirm the result due to a fortuitous HETG observation taken
nearly contemporaneously with the ACIS observation along with the RXTE
All-sky Monitor data.  GX5-1 is so bright that pileup was evident out
to a radius of $100''$.  Nonetheless, we were able to extract reliable
surface brightness profiles between $100-800''$.  GX5-1 has a large (but
rather poorly determined) column density which is likely in the range
$2-4\times10^{22}$\,cm$^{-2}$.  It is also close enough to the
Galactic plane that velocity-resolved CO measurements exist which show
the presence of a number of dense clouds along the line of sight.

We have shown that GX5-1's column density is large enough that X-rays
with energies below 2 keV are often scattered multiple times by dust
along the line of sight.  Since this is a complication to fitting, we
limited our fitting to energies above 2.5 keV where multiple
scattering is much less significant.  We compared our observations to
a wide range of dust grain models, including for the first time the
many models described in ZDA04.  We found reasonable fits for some
dust models assuming either a smooth dust distribution or a single
``cloud'' along the line of sight.  No dust grain model fit the
default parameters determined from the CO data for any distance to
GX5-1.  However, the fit parameters found when using a single cloud
suggested a re-examination of the CO data.  The 3 kpc Arm (at 5.1 kpc)
is on the boundary between the Galactic center, where the CO is
anomalously bright \citep{Sodroski95}, and the Galactic disk.  If the
gas in the 3 kpc Arm is in fact similar to the Galactic center
emission, it would have only 1/5th of the molecular gas (and
associated dust) as predicted in Table~\ref{tab:cloud}.  In this
case, most of the dust would be the one in the Molecular Ring,
which at 3.3 kpc is roughly 1/3 of the distance to the Galactic center
and which has a predicted hydrogen column density of $\sim
1.8\times10^{22}$\,cm$^{-2}$.  When using these values, we found
better fits but somewhat surprisingly still not as good as the smooth
dust distribution results.  More data from X-ray binaries at or near
the Galactic center are needed to see if in fact this interpretation
of the CO data holds when compared to other X-ray halo observations in
this area, or is merely special pleading.

We also note our strong disagreement with the results of
\citet{Xiang05} regarding the dust density along the line of sight.
Although we are not as sensitive to dust near the source as their
observation, the effects would still be noticeable in our results.
We do not know the origin of this discrepancy, but note that they
found significant amounts of dust near all of their sources.  It seems
possible that there may be calibration issues, especially in the
zero-order of the HETG, which are in some part responsible for their
result.

We also found that some dust grain models strongly disagreed with the
observations in all cases.  Most notably, nearly all of the Composite
dust models of ZDA04 (models beginning with COMP-) disagreed strongly
with the data (the sole exception is the COMP-GR-B model).  The
composite ``No Carbon'' (COMP-NC) models in particular were uniformly
poor descriptions.  The BARE-GR-B model, which combines bare graphite
and silicate grains with PAHs using B star abundances, was the overall
best fitting model in all cases.  However, since no dust model or
spatial distribution led to $\chi^2_{\nu} \sim 1$\ fits, and given
remaining uncertainties in the calibration and the modeling we can not
state that this dust model is ``preferred.''
\begin{figure}
\includegraphics[totalheight=2.3in]{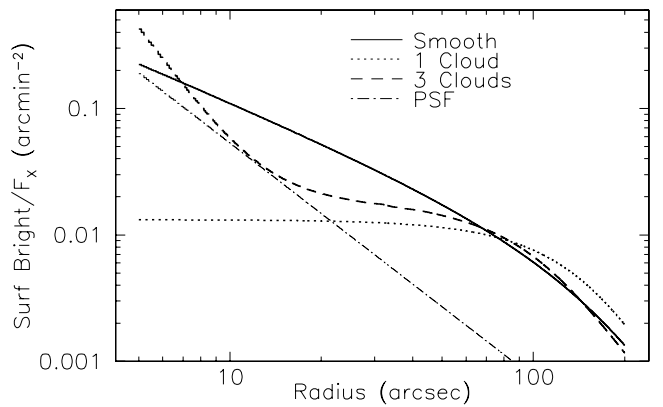}
\caption{Predicted near-source X-ray halos at 2.5 keV using the ZDA04
  BARE-GR-B model with total \NH$=3.91\times10^{22}$\,cm$^{-2}$.
  Three different dust position distributions are shown: smoothly
  distributed dust, a single cloud 1/3 of the distance to GX5-1, and
  three clouds at the positions predicted by the CO data, assuming a
  source distance of 8.51 kpc. \label{fig:NearProfile}} 
\end{figure}
The idea of directly measuring the IS dust mass, composition, and size
distribution is enticing enough that it appears in every proposal to
observe X-ray halos.  However the history of astronomical X-ray
scattering halo results shows that there are difficulties arising from
the many assumptions needed before the data can be modeled.  Although
not conclusive, our results show that adding CO and H{\sc i}\
observations do help in reducing the allowed model space and putting
better constraints on the dust models.  However, the uncertainties in
the CO and H{\sc i}\ analysis must be included to sensibly use this
additional data.  We plan to test our results by analyzing more
Galactic Center sources to see if, for example, they also imply that
the 3 kpc Arm is CO-bright.  Another way to break the existing
near-degeneracy between dust models would be to obtain measurements of
the radial profile between $10''-100''$.  Figure~\ref{fig:NearProfile}
shows that variations in dust grain positions affect the surface
brightness at these angular distances far more than in the
$100-800''$\ region measurable with our data.  We have just obtained
\chandra\ HRC-I observations of GX5-1 which will measure the radial
profile in this range (albeit without energy resolution).

\section{Acknowledgments}

We thank the anonymous referee for a detailed reading of the
manuscript which improved the paper in many ways.  In addition, we
thank Dr. Victor Zubko, Dr. Eli Dwek, and Dr. Richard Arendt for their
assistance in properly including the ZDA models.  One of the authors
(RKS) was partially supported for this work by NASA LTSA Grant
\#NNG04GC80G.


\begin{thebibliography}{}

\bibitem[Bandyopadhyay et al. (2003)]{Bandy03}  Bandyopadhyay, R.~M.,
  Shahbaz, T., \& Charles, P.~A.\ 2003, \mnras, 340, L13

\bibitem[Bania (1980)]{Bania80} Bania, T.~M.\ 1980, \apj, 242, 95 

\bibitem[Berendsen et al. (2000)]{Berendsen00} Berendsen,~S.~G.~H.,
  Fender,~R, Kuulkers,~E., Heise,~J. \& van der Klis,~M.  2000,
  \mnras, 318, 599 

\bibitem[Blitz et al. (1985)]{Blitz85} Blitz,~L., Bloemen,~J.~B.~G.~M.,
  Hermsen,~W. \& Bania,~T.~M. 1985, \aap, 143, 267

\bibitem[Bronfman et al. (1988)]{Bronfman88} Bronfman, L., Cohen, 
R.~S., Alvarez, H., May, J. \& Thaddeus, P.\ 1988, \apj, 324, 248 

\bibitem[Christian \& Swank (1997)]{Christian97} Christian,~D.~J. \&
  Swank,~J.~H. 1997, ApJSS, 109, 177 

\bibitem[Dame, Hartmann \& Thaddeus (2001)]{Dame01} Dame, T.~M., Hartmann, D. 
\& Thaddeus, P.\ 2001, \apj, 547, 792 

\bibitem[Davis (2001)]{Davis01} Davis, J.~E.\ 2001, \apj, 562, 575 

\bibitem[Dickey \& Lockman (1990)]{DL90} Dickey \& Lockman,~J. 1990,
\araa, 28, 21 

\bibitem[Draine \& Lee(1984)]{DL84} Draine, B.~T., \& Lee, 
H.~M.\ 1984, \apj, 285, 89 

\bibitem[Draine (2003)]{Draine03} Draine, B.~T.  2003, ApJ, 598, 1026

\bibitem[Guinier \& Fournet (1955)]{Guinier55} Guinier,~A. \&
  Fournet,~G.  1955, {\it Small-angle Scattering of X-rays}\ (New
  York: Wiley) 

\bibitem[Hartmann \& Burton (1997)]{Hartmann97} Hartmann, D.~\&
  Burton, W.~B.\ 1997, Cambridge; New York: Cambridge University Press

\bibitem[Henke (1981)]{Henke81} Henke, B. L. 1981, in ``Low Energy X-Ray
Diagnostics'', ed. D. T. Attwood and B. L. Henke (New York, AIP)

\bibitem[Jonker et al. (2000)]{Jonker00} Jonker,~P.~G., Fender,~R.P.,
  Hambly,~N.~C. \& van der Klis,~M.  2000, \mnras, 315, L57 \\ 

\bibitem[van der Klis (1995)]{vdK95} van der Klis, M., 1995 in Lewin
  W.H.G, van Paradijs, J, and van den Heuvel E.P.J., eds., X-ray
  Binaries (Chapter 6), Cambridge University Press, p. 252

\bibitem[Mathis, Rumpl \& Nordsieck (1977)]{MRN77} Mathis,~J.~S.,
  Rumpl,~W. \& Nordsieck,~K.~H. 1977, \apj, 280, 425 (MRN77)

\bibitem[Mathis \& Lee (1991)]{ML91} Mathis, J. S. \& Lee, C. W.
  1991, \apj, 376, 490 

\bibitem[Mauche \& Gorenstein (1986)]{MG86} Mauche, C. W. \&
  Gorenstein, P. 1986, \apj, 302, 371 

\bibitem[Overbeck (1965)]{Overbeck65} Overbeck, J. W. 1965, \apj, 141, 864

\bibitem[Predehl et al. (1992)]{Predehl92} Predehl,~P.,
  Schmitt,~J., Snowden,~S. \& Tr\"umper,~J.  1992, Science, 257, 935

\bibitem[Predehl \& Schmitt (1995)]{PS95} Predehl, P. \& Schmitt,
  J. 1995, \aap, 293, 889 

\bibitem[Predehl \& Klose (1996)]{PK96} Predehl,~P. \& Klose,~S. 1996,
  \aap, 306, 283 

\bibitem[Smith \& Dwek (1998)]{SD98} Smith,~R.~K. \& Dwek,~E.  1998,
  \apj, 503, 831 

\bibitem[Smith, Edgar \& Shafer (2002)]{SES02} Smith,~R.~K.,
  Edgar,~R.~J. \& Shafer,~R.~A.  2002, \apj, 581, 562 (SES02)

\bibitem[Sodroski et al.(1995)]{Sodroski95} Sodroski, T.~J., et 
al.\ 1995, \apj, 452, 262 

\bibitem[Ueda et al. (2005)]{Ueda05} Ueda,~Y., Mitsuda,~K,
  Murakami,~H., \& Matsushita,~K.  2005, \apj, 620, 274

\bibitem[Weingartner \& Draine (2001)]{WD01} Weingartner, J. \&
  Draine, B.\ 2001, \apj, 548, 296 (WD01)

\bibitem[Witt, Smith \& Dwek (2001)]{WSD01} Witt, A.~N., Smith, R.~K., 
\& Dwek, E.\ 2001, ApJL, 550, L201 

\bibitem[Xiang, Zhang, \& Yao (2005)]{Xiang05} Xiang,~J., Zhang,~S.~N. \&
  Yao,~Y. 2005, astro-ph/0503256 

\bibitem[Zubko, Dwek \& Arendt (2004)]{ZDA04} Zubko,~V., Dwek,~E. \&
  Arendt,~R.  2004, \apjs, 152, 211 (ZDA04) 

\end{thebibliography}
\end{document}